\documentclass[reprint,amsmath,amssymb,aps,prb,superscriptaddress,longbibliography]{revtex4-1}

\pdfoutput=1

\usepackage{hyperref,url}
\usepackage{amsmath}
\usepackage{amsfonts}
\usepackage{mathtools}
\usepackage[capitalise]{cleveref}
\usepackage{placeins}
\usepackage[version=4]{mhchem}
\hypersetup{breaklinks,colorlinks=true,linkcolor=blue,citecolor=blue,urlcolor=blue,filecolor=blue}
\usepackage{graphicx}
\usepackage{dcolumn}
\usepackage{etoolbox} 
\usepackage{float} 

\usepackage{pdfpages}
\makeatletter
\AtBeginDocument{\let\LS@rot\@undefined}
\makeatother
\usepackage{pgffor}

\newcommand\mat\mathbf

\usepackage[utf8]{inputenc}
\DeclareUnicodeCharacter{22EF}{\ldots}

\newcommand{\bytedanceus}{\affiliation{ByteDance Research, San Jose, CA 95110, US.}}
\newcommand{\bytedance}{\affiliation{ByteDance Research, Zhonghang Plaza, No. 43, North 3rd Ring West Road, 100098 Beijing, China.}}
\newcommand{\ETHZurich}{\affiliation{ETH Z{\"u}rich, R{\"a}mistrasse 101, 8092 Z{\"u}rich, Switzerland.}}

\begin{document}

\author{Hung Q. Pham}
\email{hung.pham@bytedance.com}
\bytedanceus
 
\author{Runsheng Ouyang}
\ETHZurich
 
\author{Dingshun Lv}
\email{lvdingshun@bytedance.com}
\bytedance

\title{Scalable Quantum Monte Carlo with Direct-Product Trial Wave Functions}

\begin{abstract}
The computational demand posed by applying multi-Slater determinant trials in phaseless auxiliary-field quantum Monte Carlo methods (MSD-AFQMC) is particularly significant for molecules exhibiting strong correlations. Here, we propose using direct-product wave functions as trials for MSD-AFQMC, aiming to reduce computational overhead by leveraging the compactness of multi-Slater determinant trials in direct-product form (DP-MSD). This efficiency arises when the active space can be divided into non-coupling subspaces, a condition we term ``decomposable active space''. By employing localized-active space self-consistent field wave functions as an example of such trials, we demonstrate our proposed approach across a range of molecular systems, each exhibiting varying degrees of complexity in their electronic structures. Our findings indicate that the compact DP-MSD trials can reduce computational costs substantially, by up to 36 times for the \ce{C2H6N4} molecule where the two double bonds between nitrogen \ce{N=N} are clearly separated by a \ce{C-C} single bond, while maintaining accuracy when active spaces are decomposable. In the case of larger systems such as the benzene dimer, characterized by weak coupling between the two monomers, we observed a decrease in computational cost compared to using a complete active space trial, yet we retained the same level of accuracy. However, for systems where these active subspaces strongly couple, a scenario we refer to as "strong subspace coupling", the method's accuracy decreases compared to that achieved with a complete active space approach. We anticipate that our method will be beneficial for systems with non-coupling to weakly-coupling subspaces that require local multireference treatments.
\end{abstract}
\maketitle
\newpage

\section{Introduction}

The central challenge in quantum chemistry lies in finding a computationally efficient and accurate solution to the many-electron ground-state problems.\cite{helgaker2014molecular,Helgaker2008} The difficulty stems from the fact that the most accurate methods tend to be the most computationally expensive. For large systems, it becomes necessary to employ computationally economical alternatives, such as Kohn-Sham Density Functional Theory (KS-DFT) which is known for its low computational cost and reasonable accuracy.\cite{KS-DFT,HK_theorem} KS-DFT has become the method of choice in practical applications of computational chemistry, physics, materials science, and computational biology.\cite{DFT_mat1,DFT_mat2} However, KS-DFT generally lacks predictive power, as its accuracy often depends on the choice of exchange-correlation functional, which may not be known a priori for the system of interest, and its inability to handle systems with strong electron correlation.\cite{Mardirossian2017} State-of-the-art methods beyond KS-DFT, including coupled cluster theory,\cite{Bartlett1981Oct} perturbation theory, tensor network-based methods,\cite{Orus2019} and quantum Monte Carlo (QMC),\cite{becca_sorella_2017} remain computationally expensive, even for medium-sized systems. Designing a method that is not only computationally efficient and accurate, but also capable of systematic improvement towards an exact solution, is crucial. This is pivotal for advancing the development of new drugs and functional materials, as well as gaining a deeper understanding of chemical processes that significantly contribute to technological progress.

QMC has become a widely recognized technique in quantum chemistry in recent years. Among its variants, phaseless auxiliary-field quantum Monte Carlo (ph-AFQMC) has been known for its balance between accuracy and efficiency.\cite{Zhang2003Apr,Motta2018Sep,Lee2022,Shee2023} The selection of the trial wave function, however, significantly impacts the accuracy and scalability of ph-AFQMC. The utilization of a trial wave function is essential for ensuring statistical efficiency and alleviating the renown fermionic phase (or sign) problem. When one employs multi-Slater determinant trials, constructed using a multireference method and denoted as MSD-AFQMC, this approach demonstrates remarkable accuracy for systems exhibiting strong electron correlation. It does this while maintaining a scalability that ensures reasonable computational costs.\cite{Shee2017Jun,Shee2019Sep,Shee2019Apr} Despite its demonstrated advantages, the application of ph-AFQMC with MSD trial wavefunctions can still pose a considerable computational challenge. The typical scaling is $\mathcal{O}(N_c \cdot N^4)$, where $N_c$ refers to the number of Slater determinants or configurations and $N$ represents the number of basis functions.\cite{Mahajan2021Aug} Although recent efforts have aimed to alleviate this scaling,\cite{Malone2019Jan,Motta2019Jun,Weber2022Jun,Mahajan2022May} it is important to emphasize that the number of Slater determinants~($N_c$) increases exponentially with the size of the active space. Consequently, this limits the applicability of ph-AFQMC with MSD trials to medium-sized chemical systems with small active spaces. Hence, improving the compactness of MSD trials for ph-AFQMC calculations through reducing the number of important Slater determinants $N_c$, would greatly expand the potential of ph-AFQMC for the simulation of chemical systems with complex electronic structures.

It is well recognized that there is a significant degree of sparsity in the configuration interaction wave function for a number of important real-world applications, such as the excitonic properties in molecular crystals and noncovalent interactions in biological systems.\cite{Christensen2016} This inherent sparsity, rooted in the local nature of the electron correlation, can be leveraged when generating trials for ph-AFQMC calculations to decrease computational cost. Several quantum chemical techniques have been developed for \textit{\textit{ab initio}} simulations that make use of this sparsity, for instance, selected configuration interaction~\cite{Coe2023,Holmes2016Aug} (sCI) or matrix product states methods.\cite{Frahm2019} These techniques are commonly developed as universal tools and can be applied to a broad spectrum of chemical problems, particularly those of small to moderately medium size. However, they may not inherently benefit from an intuitive understanding of the system of interest. Alternatively, there exists a variety of techniques that incorporate this understanding directly as user input. Quantum embedding methods,~\cite{Sun2016,li2022toward,Cao2023} for example, allow for the fragmentation of the system based on the local nature of the system's chemical structure. This results in powerful methods to address electron correlation in larger systems. In our pursuit of advancing chemical simulation methodologies, we aim to devise tailored and problem-specific methods suitable for large-scale chemical systems. This approach, drawing parallels to the necessity of choosing the right tool for the task in any research discipline, could prove more effective in addressing distinct challenges. Such intricacies might include understanding the binding energy and excitonic properties in molecular crystals, or probing the noncovalent interactions in biological systems. 

In this work, we introduce a novel methodology termed direct-product MSD-AFQMC, or DP-MSD-AFQMC. This approach is designed for efficient study of strongly correlated systems with decomposable active spaces, wherein a complete active space can be partitioned into non-coupling active subspaces. DP-MSD-AFQMC takes advantage of the compactness of MSD trial wave functions in a direct-product form to effectively address scalability concerns of DP-AFQMC. Our methodology accurately captures both static and dynamic correlations in these systems, while eliminating numerous unimportant configurations, owing to the inherent compactness of the trial wave functions. DP-MSD-AFQMC is effective as long as active spaces can be partitioned. We demonstrate the effectiveness of our approach through numerical experiments on various strongly correlated systems of increasing complexity. Via these experiments, we offer guidance on the types of systems for which DP-MSD-AFQMC is ideally suited, as well as those for which it may not be the preferred choice.

\section{Theory}\label{sec:theory}
\subsection{Overview of ph-AFQMC: Theory and Formal Scaling}\label{sec:afqmc}
We provide a short overview of ph-AFQMC theory to establish the background necessary for discussing the computational challenges associated with ph-AFQMC for strongly correlated molecules. While there is extensive literature on the implementation details and phaseless approximation of ph-AFQMC, we will only present the relevant equations and assumptions that support our discussion of the key challenges addressed in this study. Readers can refer to other comprehensive reviews for more information.\cite{Motta2018Sep,Lee2022}

The formulation of ph-AFQMC typically begins with the process of imaginary-time propagation of an initial wave function.

\begin{equation}
|\Psi\rangle = \lim_{\tau \rightarrow \infty} e^{-\tau \hat{H}} |\Phi_0\rangle
\end{equation}

where $|\Phi_0\rangle$ is an initial wave function that has some overlap with $|\Psi\rangle$ (which is the unknown ideal ground state), $\tau$ is the imaginary time, and $\hat{H}$ is the Hamiltonian. This equation constitutes the theoretical foundation behind any projector QMC methods, with the ground state being an asymptotic solution of the imaginary-time Schrödinger equation.

The ph-AFQMC algorithm employs an importance sampling technique that is based on a trial wave function, denoted as $|\Psi_T\rangle$ during the open-ended random walk process. The global wave function at a given time $\tau$ can be written as a weighted statistical sum over $N_\omega$ walkers $|\psi_i(\tau)\rangle\rangle$

\begin{equation}
|\Psi(\tau)\rangle = \sum_{i=1}^{N_\omega} \frac{\omega_i(\tau)}{\langle \Psi_T |\psi_i(\tau)\rangle} |\psi_i(\tau)\rangle
\end{equation}

where $\omega_i(\tau)$ is the walker weight and $|\Psi_T\rangle$ is the trial wave function. The local energy, $E_{loc}$, for each walker at a given time $\tau$ can be estimated using the mixed estimator.

\begin{equation}
E_{loc} = \frac{\langle \Psi_T|\hat{H}|\psi_i(\tau)\rangle}{\langle \Psi_T|\psi_i(\tau)\rangle}
\end{equation}

The global energy at any time $\tau$ can be estimated in a statistical manner.

\begin{equation}
E_0(\tau) = \frac{\sum_{i=1}^{N_\omega} \omega_i(\tau)E_{loc}}{\sum_{i=1}^{N_\omega} \omega_i(\tau)}
\end{equation}

Next, we provide only a summary of the main steps and their associated scaling, with the aim of establishing a foundation for the discussion of how using direct-product multi-Slater determinant trial can reduce the computational cost of ph-AFQMC calculations. We note that further advanced techniques tailored for each component in the scaling analysis can be incorporated into our current methods, without affecting the final conclusions drawn here. When assessing the computational cost of a ph-AFQMC calculation, there are three key components to consider:

(1) Obtain the overlap between the trial wave function and the walker, $\langle \Psi_T|\psi_i(\tau)\rangle$.

(2) Evaluate the Green's function for calculating the force bias as described in more detail in reference \cite{Lee2022}.

(3) Estimate the local energy $E_{loc}$. The scaling for computing the local energy, as presented in Table~\ref{table:scaling}, follows a standard quartic scaling. However, more recent advanced algorithms have the potential to reduce this to cubic scaling through methods such as double factorization,~\cite{Motta2019Jun} tensorhypercontraction,~\cite{Malone2019Jan}, stochastic resolution-of-the-identity,~\cite{Lee2020Jul} or by utilizing localized orbitals.~\cite{Weber2022Jun}

\begin{table}[h]
\centering
\caption{The computational cost per sample for overlap, Green’s function, and local energy. $N$ and $N_c$ denotes the number of basis functions and the number of determinants, respectively.}
\begin{tabular}{l c}
\hline
Method & Cost per sample \\
\hline
Overlap & $\mathcal{O}(N_c + N^3)$~\cite{Mahajan2021Aug} \\
Green's function & $\mathcal{O}(N_c + N^3)$~\cite{Mahajan2022May} \\
Local energy & $\mathcal{O}(N_c \cdot N^4)$ \\
& $\mathcal{O}(N_c \cdot N + N^4)$~\cite{Mahajan2021Aug} \\
& $\mathcal{O}(N_c \cdot N^2 + N^3)$~\cite{Weber2022Jun} \\
\hline
\label{table:scaling}
\end{tabular}
\end{table}

Table~\ref{table:scaling} displays the computational complexity associated with evaluating the local energy in ph-AFQMC calculations using MSD trials. We note that the number of determinants in a CI wave function has a subtle dependence on the number of basis sets. The cost analysis for stochastic methods, such as AFQMC, is nuanced, making direct comparisons with deterministic methods less straightforward. It's often recommended to estimate the cost based on a fixed statistical error. Although advanced update algorithms like fast updates~\cite{Shi2021Jan} can mitigate the scaling to sublinear in $N_c$, this complexity becomes especially significant for molecular systems that are extensive in size and necessitate a considerable number of determinants. To alleviate this complexity, reducing the number of determinants $N_c$ in the MSD trials can significantly improve the computational cost of ph-AFQMC for extended molecules. Our proposed approach leverages the compactness of wave functions that contain weakly coupled active spaces, which is discussed in further detail in the following section.

\subsection{Direct-product trial wave function}\label{sec:dp-msd}

Without loss of generality, we assume the system is composed of two non-coupling active subspace $A$ and $B$, respectively. 
Thus, the ground state wave function $|\Psi\rangle$ can be described by a direct-product form of configuration interaction (CI) expansion:

\begin{equation}
|\Psi\rangle = \sum_{I,J} C_{IJ}|\Phi_I^A\rangle \otimes |\Phi_J^B\rangle = \sum_{I,J} C_{IJ}|\Phi_I^A \Phi_J^B\rangle
\end{equation}

where $|\Phi_I^A\rangle$ and $|\Phi_J^B\rangle$ denote the many-electron basis states, commonly referred to as configurations, for active subspaces $A$ and $B$, respectively; $C_{IJ}$ is the coefficient of the CI expansion. The above configuration interaction (CI) expansion for the composite system implicitly fulfills the requirement for the Pauli antisymmetry of fermions. It is worth mentioning that the number of configurations grows exponentially with the size of the combined active space, making direct diagonalization of the Hamiltonian infeasible for larger systems due to prohibitive computational demands.

The key to efficiently computing the above ground state wave function lies in the ability to diagonalize the Hamiltonian without explicitly constructing the direct-product basis state $|\Phi_I^A \Phi_J^B\rangle$. The Hamiltonian corresponding to each active space can be diagonalized independently and expressed as follows:

\begin{align}
|\Psi_A\rangle &= \sum_{I} C_{I}|\Phi_I^A\rangle \\
|\Psi_B\rangle &= \sum_{J} C_{J}|\Phi_J^B\rangle
\end{align}

where $C_{I}$ and $C_{J}$ are the coefficient of the CI expansion for active spaces $A$ and $B$, respectively. This yields a significant result: the number of required intermediates for computing the wave function scales linearly with the number of active subspaces. Regarding the memory requirements, while the storage needed for each CI vector in each active space increases exponentially with the size of active spaces $A$ or $B$, the total storage cost for the combined system is simply the sum of these two costs. Consequently, this approach presents a contrast to directly diagonalizing the total wave function, where the computational cost swiftly becomes infeasible due to the exponential increase.

The composite CI coefficients can be estimated from the tensor product of the monomer CI vectors:

\begin{equation}
C_{IJ} \approx C_I \otimes C_J
\end{equation}

The tensor $C_{IJ}$ within the direct-product approximation of the wave function exhibits inherent sparsity, which allows the elimination of numerous configurations with near-zero coefficients from the CI expansion. This is achievable by setting a user-defined cutoff value, $\epsilon$, for the magnitude of the coefficients. During practical calculations, the cutoff value is gradually decreased to incorporate more configurations and achieve convergence of the truncated trial energy towards the energy obtained from the CI expansion trial that utilizes all available configurations. By balancing computational feasibility with the pursuit of convergence in accuracy, our approach seeks to achieve a compromise between these two objectives. The resulting wave function can be represented by the following equation, which defines a direct-product multi-Slater determinant trial, $|\Psi_0^\epsilon\rangle$:

\begin{equation}
|\Psi_0\rangle \approx |\Psi_0^\epsilon\rangle = \sum_{I,J} C_{IJ}|\Phi_I^A \Phi_J^B\rangle \quad \text{such that} \quad |C_{IJ}| > \epsilon
\end{equation}

The memory requirement for storing the wave function, $|\Psi_0^\epsilon\rangle$, is significantly lower compared to storing all configurations. Our approach adopts a truncation procedure similar to that utilized in the sCI methods, where the CI expansion of the variational wave function is sorted and truncated to include only the most relevant configurations.

Several algorithms have been proposed for generating direct-product wave functions for \textit{ab initio} Hamiltonians, including active space decomposition (ASD),\cite{Parker2013} low-rank approximation (LRA) using rank-one basis states,\cite{Nishio2019} and localized-active space self-consistent field (LASSCF).\cite{Hermes2019,Hermes2020} At their core, these methods offer an approximation for the wave function of the complete active space self-consistent field (CASSCF) method,~\cite{Siegbahn1980,Siegbahn1981,Roos1980} which can, at its limit, handle 18 electrons in 18 orbitals. Depending on the chosen direct-product wave function, the resulting method is referred to as ASD-AFQMC, LRA-AFQMC, or LAS-AFQMC, and they all fall under the larger umbrella of direct-product multi-Slater determinant ph-AFQMC or DP-MSD-AFQMC. This notation follows the common convention in quantum chemistry of a ``static then dynamic'' treatment for electron correlation, which is conveniently decomposed into dynamic and static components. Although the distinction between dynamic and static correlation within the ph-AFQMC framework is not always clear, this convention is adopted for convenience. In our numerical simulations, we choose to primarily utilize the LASSCF wave function to demonstrate the effectiveness of our method, as generating a trial wave function from LASSCF is straightforward and convenient.

\begin{figure*}[!ht]
    \centering
    \scalebox{1.0}{\includegraphics{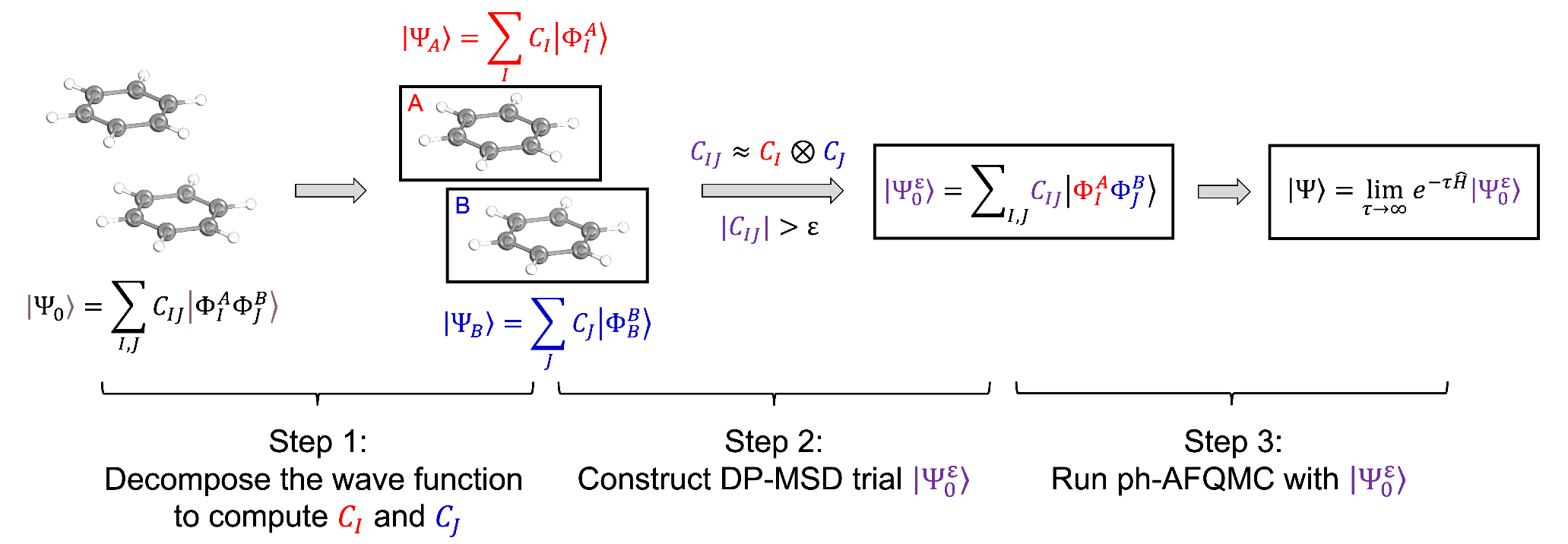}}
    \caption{Flowchart for running ph-AFQMC with a DP-MSD trial on the benzene dimer.}
    \label{fig:C2H4N4-pec}
\end{figure*}

\section{Computational details}\label{sec:computational}

The geometries for \ce{C2H6N4}, \ce{C2H4N4}, and the benzene dimer at equilibrium separation were taken from literature.\cite{Hermes2019,Herman2023} The calculations for \ce{C2H6N4} and \ce{C2H4N4} were performed using the 6-31G basis set while STO-6G and 6-31G basis set were employed for the benzene dimer. For the \ce{H10} molecule, the STO-6G and cc-pvDZ basis sets were used. In ph-AFQMC calculations, we employed 640 walkers for all the systems. For the parallel-displaced benzene dimer calculated using the 6-31G basis set with an $\epsilon$ value of $5 \times 10^{-2}$, increasing the number of walkers to 1320 doesn't significantly alter the energy as shown in Table S7 of the of the Supporting Information. Consequently, we believe that this number of walkers is adequate for other smaller systems investigated in this study. A time step ($\Delta\tau$) of 0.005 a.u. was used for all structures, with the exception of the benzene dimer at equilibrium separation, for which an additional time step of 0.001 a.u. was also applied. This particular case is elaborated upon in the discussion section. The trial wave functions were generated using an in-house code. The details of the active spaces used in the CASSCF and LASSCF calculations are presented in Figure S1-S3 in the Supporting Information. QMCPACK was used for the ph-AFQMC calculations.\cite{Kent2020May} The LASSCF wave function was generated using the mrh code,\cite{mrh} while integrals and mean-field calculations, as well as CASSCF calculations, were performed using PySCF.\cite{Sun2020Jul} The MRCI+Q calculations were carried out using BAGEL,\cite{BAGEL} and semistochastic heat-bath configuration interaction~(SHCI) calculations were performed using Dice.~\cite{Holmes2016, Sharma2017}

\section{Results}\label{sec:results}

\subsection{\ce{C2H6N4} and \ce{C2H4N4}}\label{sec:bisdiazenes}

We first investigate the \ce{C2H6N4} molecule where the two double bonds between nitrogen \ce{N=N} are clearly separated by a \ce{C-C} single bond. Previous studies have shown that this molecule is well-suited for decomposing the total active space of eight electrons in eight $\pi$ orbitals or CAS(8,8) into the product of two active spaces of four electrons in four $\pi$ orbitals or CAS(4,4).

\begin{figure*}[!ht]
    \centering
    \scalebox{1.}{\includegraphics{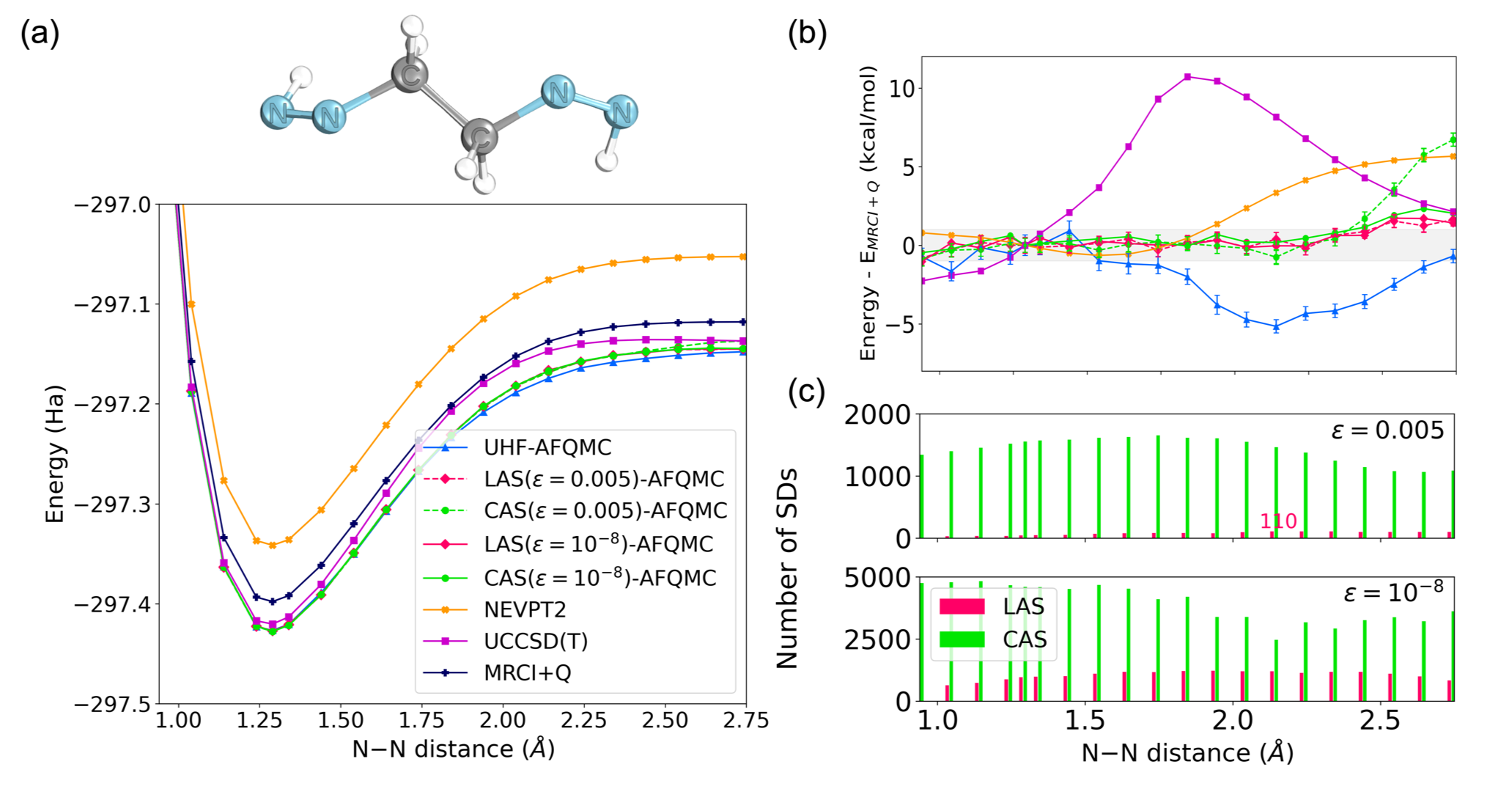}}
    \caption{(a) Potential energy curves of \ce{C2H6N4} obtained using different methodologies. (b) Energy deviations from MRCI+Q for each method, with the errors at the equilibrium distance shifted to zero. (c) Number of determinants as a function of \ce{N-N} distance for two different cutoff values: $\epsilon= 0.005$ and $\epsilon=10^{-8}$. The grey panel indicates the chemical accuracy of 1 kcal/mol. The colors of the balls in the figures represent different atoms: blue for nitrogen, grey for carbon, and white for hydrogen.}
    \label{fig:C2H6N4}
\end{figure*}

In Figure~\ref{fig:C2H6N4}(a), we present potential energy curves generated by various methodologies, including AFQMC with different types of trials, N-electron valence state perturbation theory (NEVPT2), unrestricted coupled cluster with Single and Double excitations (UCCSD(T)), and multireference configuration interaction method plus Davidson correction (MRCI+Q). The MRCI+Q method is prevalent in the literature because of its high accuracy for strongly correlated systems, and in this study, it serves as the benchmark for all other methods. Our MRCI+Q calculations employ the same CASSCF wave function, ensuring they possess the identical number of determinants as the CAS trial. The deviation plots for the different methodologies are illustrated in Figure~\ref{fig:C2H6N4}(b), with all curves being shifted to zero at the equilibrium bond length of 1.29 Å. Deviation plots are better suited for comparing different quantum chemical methods, particularly when examining chemical reaction barriers. We observe that ph-AFQMC using a single-Slater determinant trial like unrestricted Hartree-Fock (UHF) (indicated by the blue curve) is inaccurate in the N-N distance range between approximately 1.50 and 2.70 Å, but performs well for other distances. Similarly, AFQMC using CASSCF as the trial, with $\epsilon= 0.005 $ as the cutoff, denoted by the dashed green curve, is highly accurate up to about 2.30 Å, beyond which its accuracy rapidly declines. However, this can be remedied by employing a much tighter cutoff of $\epsilon=10^{-8}$, denoted by the solid green curve. Surprisingly, AFQMC using the trials constructed from a LASSCF wave function are the most accurate across all bond lengths, regardless of the cutoff values used. This demonstrates the compactness of the direct-product multi-Slater determinant trial as discussed in the theoretical analysis. Additionally, it is important to note that the common NEVPT2 technique is unable to accurately calculate the potential energy curves of this molecule, with deviations as large as 5 kcal/mol observed at the strong correlation limit.

\begin{figure}[!ht]
    \centering
    \scalebox{0.55}{\includegraphics{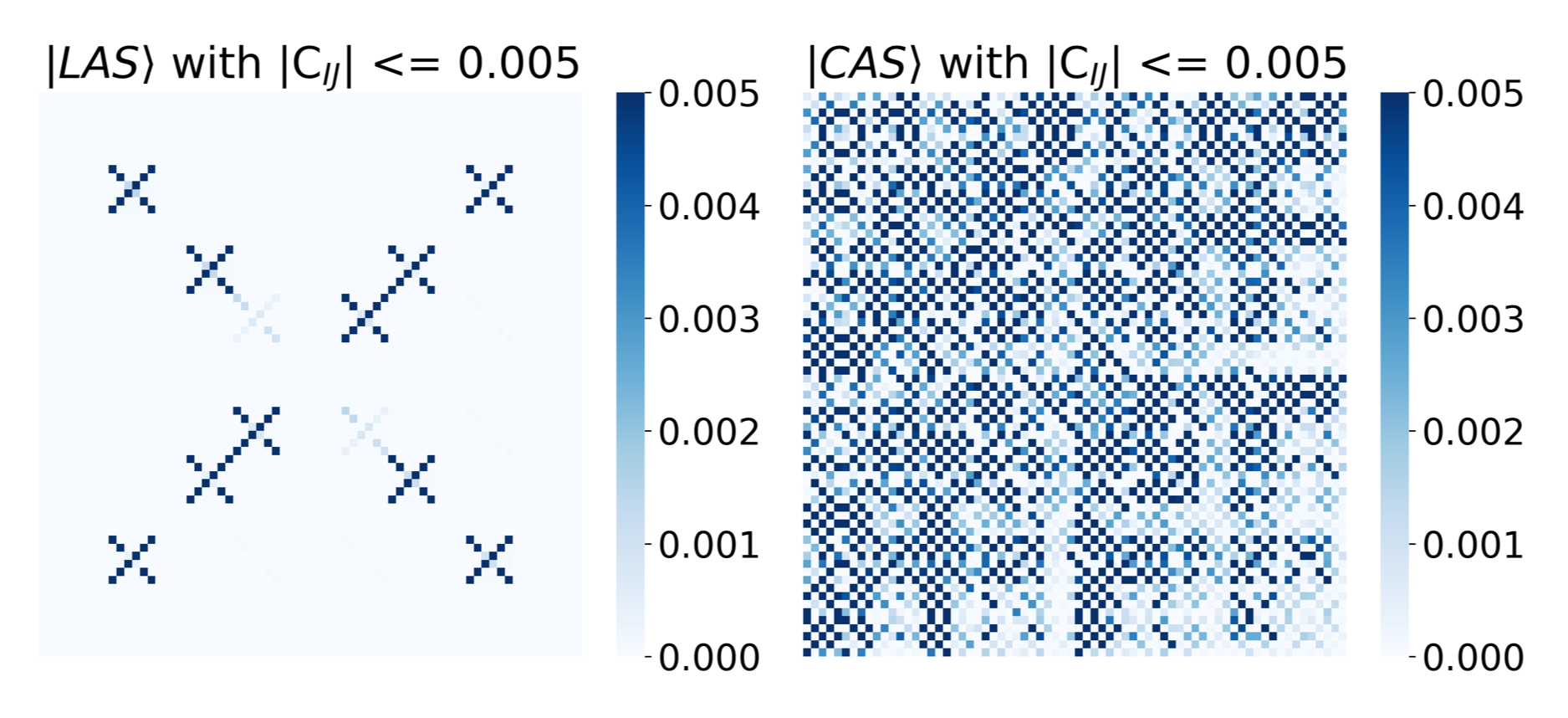}}
    \caption{Heatmaps illustrating the CI tensor for the LAS and CAS trials at $\epsilon= 0.005 $.}
    \label{fig:C2H6N4-CI}
\end{figure}

To evaluate the computational efficiency of our method, we plot the number of determinants as a function of the \ce{N-N} distance for two cutoff values, namely $\epsilon= 0.005 $ and $\epsilon=10^{-8}$, for both CASSCF and LASSCF wave functions. Figure~\ref{fig:C2H6N4}(c) shows that the number of Slater determinants in the LASSCF trial is considerably smaller than that in the CAS trial when $\epsilon= 0.005$  is used, compared to when $\epsilon=10^{-8}$ is used, confirming the sparsity of the LASSCF wave function and of the direct-product wave function in general.Furthermore, our findings reveal that the LAS($\epsilon= 0.005 $)-AFQMC method, with CAS-AFQMC as the reference, exhibits the highest accuracy among the investigated methods, requiring a maximum of 110 Slater determinants to study the potential energy curve (PEC) of \ce{C2H6N4}. In contrast, despite utilizing over 1000 Slater determinants, the CAS($\epsilon = 0.005$)-AFQMC method fails to accurately calculate the PEC at the strong correlation limit. Notably, CAS($\epsilon=10^{-8}$)-AFQMC necessitates up to 4000 determinants to achieve a comparable level of accuracy. This results in a remarkable computational cost reduction of approximately 36 times (4000 divided by 110). Upon examining the CI expansion of each trial, it's clear that the main distinction between the LAS and CAS trials for this system is in the type of determinants, depicted in Figure~\ref{fig:C2H6N4-CI}, which presents a heatmap plot for the CI tensors at $\epsilon= 0.005$. The LAS, intentionally, omits the unimportant SDs accounting for the weak coupling between the two \ce{N=N} double bonds in \ce{C2H6N4}, aligning with the sparse CI tensor. In contrast, the CAS trial incorporates these SDs, which contribute notably to the CI expansion. When these SDs are truncated, the truncated CAS trial becomes inferior to the LAS trial.

\begin{figure}[!ht]
    \centering
    \scalebox{0.6}{\includegraphics{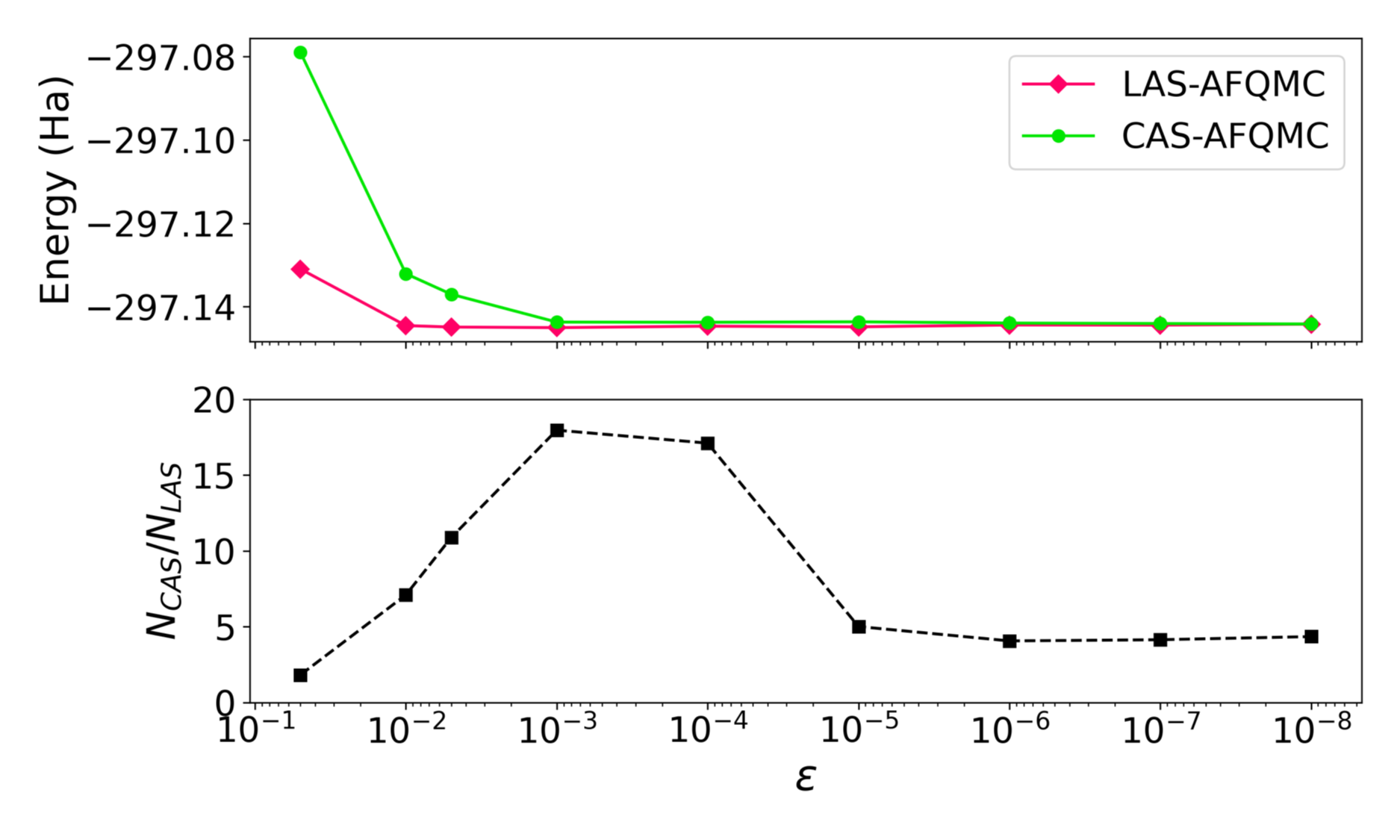}}
    \caption{Upper panel: Convergence of total energy for \ce{C2H6N4} with a \ce{N=N} bond length of approximately 2.74 Å, plotted against $\epsilon$. Lower panel: Ratio of the number of determinants in the CAS ($L_{CAS}$) to those in the LAS trial ($N_{LAS}$).}
    \label{fig:C2H6N4_scan}
\end{figure}

Additionally, we also plot the MSD-AFQMC total energy at the strong correlation limit (approximately 2.74 Å) with respect to different values of cutoff (refer to Figure~\ref{fig:C2H6N4_scan}). Our observations highlight that the LAS trial converges rapidly at $\epsilon= 0.001 $, which is equivalent to 100 SDs. In contrast, the CAS trial converges more slowly, necessitating $\epsilon= 0.001 $ (equivalent to 2012 SDs) to achieve the same accuracy within chemical accuracy. The significant reduction in the number of Slater determinants used in LAS-AFQMC supports the notion that this trial, as well as DP-MSD trials more generally, is not only accurate but also highly efficient for systems where the complete active space can be decomposed.

Next, we investigate a more challenging case for the direct-product \textit{ansatz}: \ce{C2H4N4}, where the nitrogen double bonds are coupled to each other through the \ce{C=C} double bond. Due to the presence of this double bond, two plausible active spaces for the system are CAS(10,10), where all the $\pi$ orbitals are included in the active space, and CAS(8,8), where only the $\pi$ orbitals that involve nitrogen atoms are included. The corresponding active-space decompositions for these are LAS(8,8), where the CAS(8,8) is split into two CAS(4,4) spaces, and LAS(10,10), where the CAS(10,10) is split into two CAS(4,4) spaces and one CAS(2,2) space corresponding to the $\pi$ orbitals that involve nitrogen atoms nitrogen and carbon atoms, respectively. We use a tight cutoff of $\epsilon=10^{-8}$ for the DP-MSD trials. Our results indicate that while NEVPT2 or MRCI+Q are highly sensitive to the selected active spaces, the differences obtained using ph-AFQMC are relatively small, as demonstrated in Figure~S4 in the Supporting Information. Therefore, we only present the results obtained using CAS(10,10) or LAS(10,10) in the subsequent discussion.

\begin{figure}[!ht]
    \centering
    \scalebox{0.5}{\includegraphics{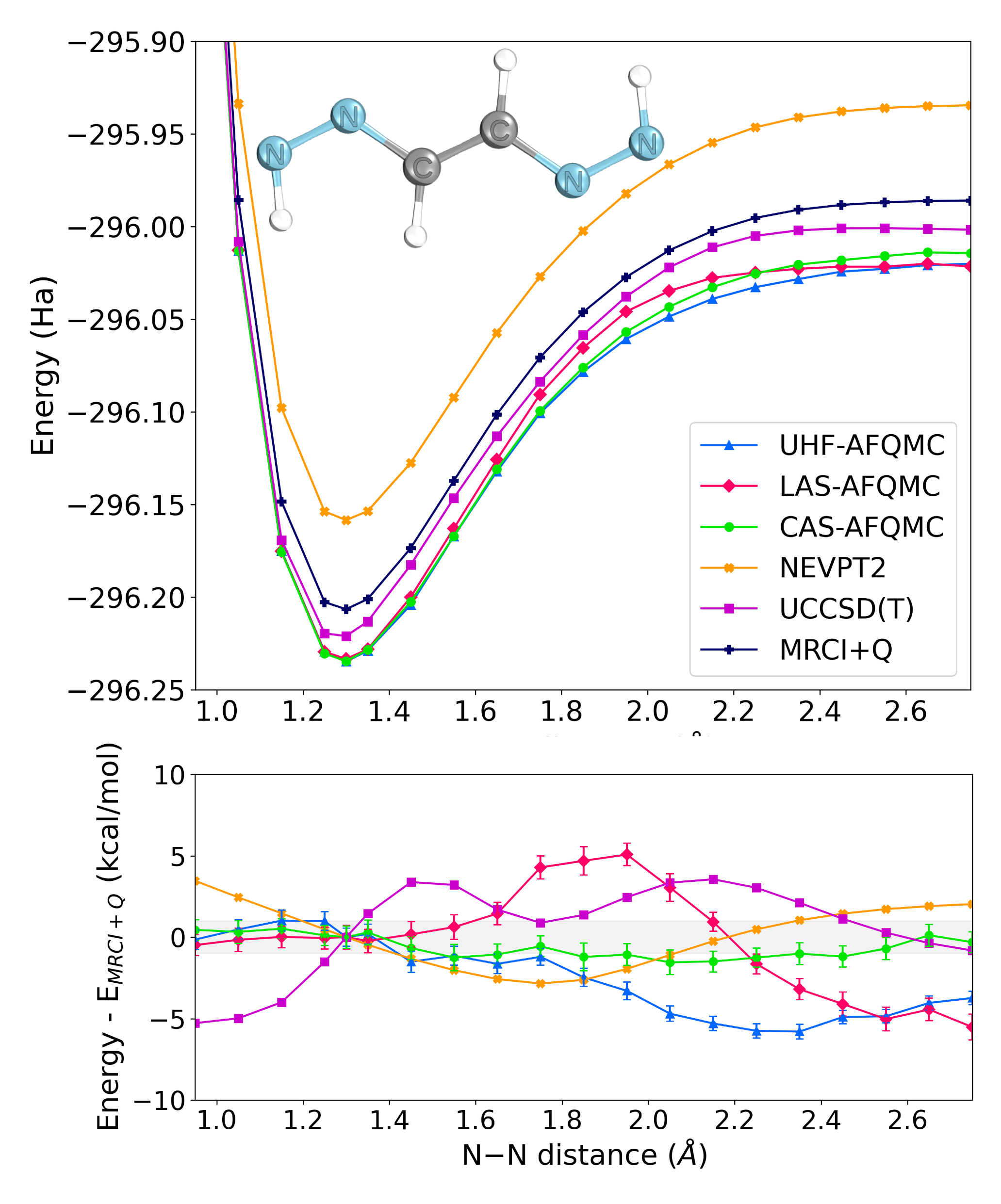}}
    \caption{Upper panel: Potential energy curves of \ce{C2H4N4} obtained using different methodologies. Lower panel: Energy deviations from MRCI+Q for each method, with the errors at the equilibrium distance shifted to zero. The grey panel indicates the chemical accuracy of 1 kcal/mol. The colors of the balls in the figures represent different atoms: blue for nitrogen, grey for carbon, and white for hydrogen.}
    \label{fig:C2H4N4}
\end{figure}

Figure~\ref{fig:C2H4N4} presents the potential energy curves of \ce{C2H4N4} obtained using various methodologies, along with their corresponding deviations from MRCI+Q. Although CAS-AFQMC yields excellent agreement at all \ce{N-N} distances, LAS-AFQMC struggles to capture the correct curvature of the potential energy curve beyond the equilibrium distance. Interestingly, LAS-AFQMC performs no better than the UHF-AFQMC, NEVPT2, or UCCSD(T) methods. These observations suggest that the active-space decompositions used for strongly coupled active spaces may inherently remove the correct physics from the wave function. Furthermore, the phaseless approximation used in the AFQMC method appears unable to recover these unphysical decompositions. 

\subsection{Hydrogen ring}\label{sec:h10}

Here, we investigate the strongly correlated toy model \ce{H10}, which is commonly used in the literature to benchmark new numerical methods due to its rich physics and simplicity. At the minimal basis set, \ce{H10} shares similar electronic structure with the half-filled Hubbard model, except that long-range interactions are included. For this system, the plausible active space is CAS(10,10), which includes two electrons in two $\sigma$ and $\sigma^*$ orbitals for each \ce{H-H} bond. In the LASSCF calculations, this active space is decomposed into five CAS(2,2) subspaces. It is worth noting that within the STO-6G basis, CAS(10,10) corresponds essentially to a full configuration interaction~(FCI) wave function.

\begin{figure*}[!ht]
    \centering
    \scalebox{1.0}{\includegraphics{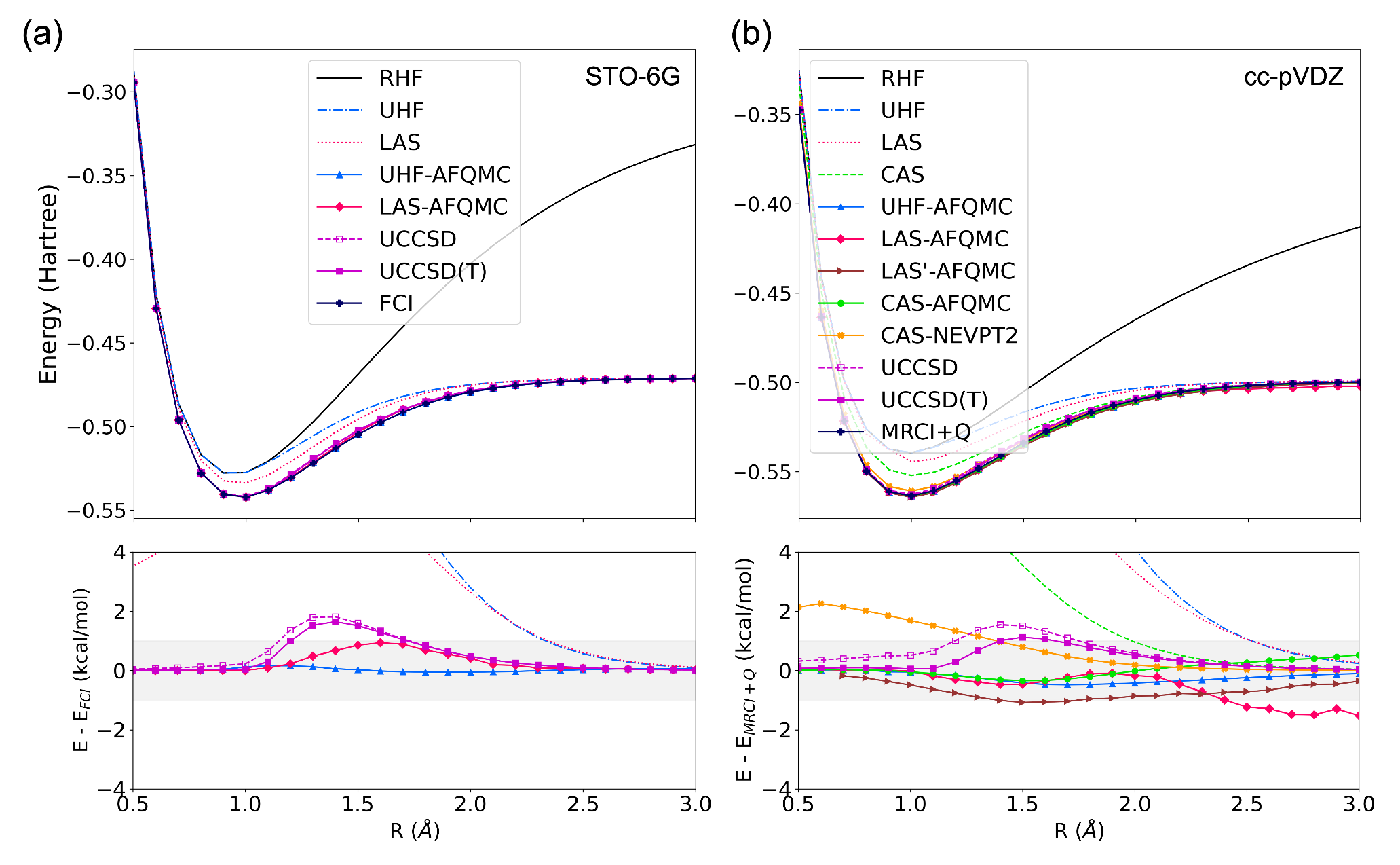}}
    \caption{Potential energy curves from the equidistant stretching of \ce{H10} obtained using different methodologies are shown in the top panels, while the corresponding deviations from FCI or MRCI+Q are displayed in the bottom panels. The right panels (a) represent results obtained using the STO-6G basis set, while the left panels (b) use cc-pVDZ. R is the distance between two neighboring hydrogen atoms}
    \label{fig:H10}
\end{figure*}

At the STO-6G basis set, we find that both CAS-AFQMC and LAS-AFQMC agree well with FCI within the chemical accuracy as shown in Figure~\ref{fig:H10}. Notably, LAS-AFQMC performs better than UCCSD(T) for bond lengths around 1.5 \r{A}. However, within the cc-pVDZ basis set, LAS-AFQMC is accurate only up to a bond length of 2.3 \r{A}, beyond which the deviation from MRCI+Q increases with the largest deviation being about 1.9 kcal/mol. This behavior is similar to the unphysical decomposition observed in the case of the \ce{C2H4N4} molecule where all five single \ce{H-H} bonds are strongly coupled and cannot be separated. In an effort to remedy this issue, we explore an alternative way to partition the system by decomposing CAS(10,10) into two CAS(5,5) states with opposite $S_z$ values that can couple to create an $\langle S^2\rangle=0$ system. Interestingly, the AFQMC calculation using this approach, denoted as LAS'-AFQMC in Figure~\ref{fig:H10}, agrees with MRCI+Q within the chemical accuracy at all bond lengths. This suggests that a direct-product trial combined with a suitable fragmentation scheme can provide a tailored solution to the problem of interest with some heuristics from chemical intuition to guide the exploration. Interestingly, the simple UHF trial performs just fine for \ce{H10}, making it a better choice than other MSD trials for this system.

\subsection{Benzene dimer}\label{sec:dibenzene}
Having established the effectiveness of DP-MSD-AFQMC for simple systems, we now focus on demonstrating the effectiveness of this method for a larger system, \textit{i.e.}, a benzene dimer which is composed of two benzene molecules interacting via nonbonding $\pi-\pi$ interactions.\cite{Herman2023} Specifically, we compute the frozen-core correlation energy for the parallel-displaced or PD ($C_{2h}$) configuration of the benzene dimer at STO-6G and 6-31G basis set, motivated by previous collaborative efforts on the benzene molecule.\cite{Eriksen2020} These efforts employed state-of-the-art numerical approximations to determine the exact frozen-core correlation energy of the system at cc-pVDZ basis set. A subsequent study highlighted the importance of utilizing MSD trials based on CAS(6,6) to reach chemical accuracy.\cite{Lee2020May} The benzene dimer presents a more challenging computational task due to its complete active space being twice the size of that for the benzene molecule. However, the nonbonding interaction between the two molecules allows us to decompose the active subspaces localized at each benzene molecule. This characteristic makes it an ideal test case for our DP-MSD-AFQMC technique. Specifically, we split the CAS(12,12) active space into two CAS(6,6) active spaces within the LASSCF trial. In our calculations, the inner core $1s$ orbitals of carbon atoms are maintained as frozen, yielding a correlation space of 60 electrons in 60 orbitals with the STO-6G basis and 60 electrons in 120 orbitals with the 6-31G basis. Additionally, we investigate the impact of various cutoff values on the convergence of the ph-AFQMC energy concerning the number of Slater determinants. We further delve into examining the influence of the strength of the coupling between two benzene molecules on the performance of the different trials discussed here. By separating the two benzene molecules from their equivalent distance of 3.48 Å to a separation of 5 Å, the $\pi-\pi$ interaction weakens, thereby enhancing the decoupling between the two molecules. This specific separation configuration is defined as PD5. Hypothetically, this becomes a more ideal scenario for the application of the LAS-AFQMC.

\begin{table*}[tp] 
\centering
\caption{Frozen-core total (in Ha) and correlation energy (in mHa) at the STO-6G level for the parallel-displaced benzene dimer at both equilibrium (PD) and 5 Å separated geometries (PD5). The correlation energy is defined as the difference between ph-AFQMC and RHF values.}
\begin{tabular}{l l c c c c}
\hline
Configuration & Method & $\epsilon$ & $N_c$ & Total energy (Ha) & Correlation energy (mHa) \\
\hline
PD & RHF  &  &  & -460.258079 & 0.0 \\
 & LASSCF &  &  & -461.482981 & -1224.9 \\
 & CASSCF &  &  & -461.483393 &  -1225.3 \\
 & RHF-AFQMC & & 1 & -461.12927 $\pm$ 0.00063 & -871.2 $\pm$ 0.6\\
 & LAS-AFQMC & $10^{-3}$ & 959 & -462.08204 $\pm$ 0.00044 & -1824.0 $\pm$ 0.4  \\
 & CAS-AFQMC & $10^{-3}$ & 3589 & -462.08229 $\pm$ 0.00045 & -1824.2 $\pm$ 0.5 \\
\hline
PD5 & RHF  &  &  & -460.260076 & 0.0 \\
 & LASSCF &  &  & -461.436737 & -1176.7 \\
 & CASSCF &  &  & -461.436794 & -1176.7 \\
 & RHF-AFQMC & & 1 & -461.12843 $\pm$ 0.00060  & -868.4 $\pm$ 0.6\\
 & LAS-AFQMC & $10^{-3}$ & 959 & -462.03572 $\pm$ 0.00028 & -1775.6 $\pm$ 0.3 \\
 & CAS-AFQMC & $10^{-3}$ & 3533 & -462.03539 $\pm$ 0.00026 & -1775.3 $\pm$ 0.3 \\
\hline
\label{table:PD-STO6G}
\end{tabular}
\end{table*}

In our initial analysis, we focus on results derived from STO-6G calculations at a cutoff value of $\epsilon=10^{-3}$. This particular cutoff was selected due to the observed convergence of total energies across all examined cases. As illustrated in Table.~\ref{table:PD-STO6G}, there's a remarkable consistency between LAS-AFQMC and CAS-AFQMC concerning the frozen-core correlation energy across both configurations. Notably, both MSD-AFQMC approaches retrieve a considerably greater correlation energy than an RHF trial, underscoring the importance of deploying an MSD trial for these systems. Significantly, LAS-AFQMC accomplishes this precision with only around 950 determinants, in contrast to the approximately 3500 Slater determinants that CAS-AFQMC necessitates. It should also be highlighted that even at a cutoff of $\epsilon = 5 \times 10^{-2}$, the deviation in total energy from the $\epsilon = 10^{-3}$ value is a mere 2-3 mHa, as presented in Tables S1 and S3 of the Supplementary Information. This slight variation can be attributed to the absence of diffuse and polarized basis functions in the STO-6G basis, resulting in artificially compact trial functions. Following this, we will investigate the slightly larger 6-31G basis set to determine if the performance consistency of the trials persists.

\begin{table*}[tp] 
\centering
\caption{Frozen-core total (in Ha) and correlation energy (in mHa) at the 6-31G level for the parallel-displaced benzene dimer at both equilibrium (PD) and 5 Å separated geometries (PD5). The correlation energy is defined as the difference between ph-AFQMC and RHF values.}
\begin{tabular}{c l c c c c}
\hline
 & Method & $\epsilon$ & $N_c$ & Total energy (Ha) & Correlation energy (mHa) \\
\hline
PD & RHF  &  &  & -461.242602 & 0.0 \\	
$\Delta\tau = 0.005$& LASSCF &  &  & -461.991116 & -748.5 \\	
 & LASSCF &  &  & -461.991850 & -749.2 \\	
 & RHF-AFQMC &  &  & -462.41604 $\pm$ 0.00060 & -1173.4 $\pm$ 0.6\\
 & LAS-AFQMC & $10^{-3}$ & 1239 & -462.99620 $\pm$ 0.00162 & -1753.6 $\pm$ 1.6 \\	
 & CAS-AFQMC & $10^{-3}$ & 4721 & -462.99579 $\pm$ 0.00183 & -1753.2 $\pm$ 1.8 \\	
 & LAS-AFQMC & $10^{-5}$ & 12684 & -462.99605 $\pm$ 0.00157 & -1753.5 $\pm$ 1.6 \\	
 & CAS-AFQMC & $10^{-5}$ & 106501 & -462.99478 $\pm$ 0.00141 & -1752.2 $\pm$ 1.4 \\	
\hline
PD & LAS-AFQMC & $10^{-3}$ & 1239 & -462.99379 $\pm$ 0.00166 &  -1751.2 $\pm$  1.7 \\	
$\Delta\tau = 0.001$& CAS-AFQMC & $10^{-3}$ & 4721 & -462.99353 $\pm$ 0.00161 &  -1750.9 $\pm$ 1.6 \\	
 & LAS-AFQMC & $10^{-5}$ & 12684 & -462.99571 $\pm$ 0.00181 &  -1753.1 $\pm$ 1.8 \\	
 & CAS-AFQMC & $10^{-5}$ & 106501 & -462.99499 $\pm$ 0.00171 &  -1752.4 $\pm$ 1.7 \\	
\hline
PD5 & RHF  &  &  & -461.246399	& 0.0 \\	
$\Delta\tau = 0.005$ & LASSCF &  &  & -461.883595 & -637.2 \\
 & CASSCF &  &  & -461.882805 & -636.4 \\
 & RHF-AFQMC &  &  & -462.41398 $\pm$ 0.00062 & -1167.6 $\pm$ 0.6\\
 & LAS-AFQMC & $10^{-3}$ & 1259 & -462.88264 $\pm$ 0.00096 & -1636.2 $\pm$ 1.0\\
 & CAS-AFQMC & $10^{-3}$ & 4909 & -462.88267 $\pm$ 0.00091 & -1636.3 $\pm$ 0.9 \\
 & LAS-AFQMC & $10^{-5}$ & 9024 & -462.88204 $\pm$ 0.00114 & -1635.6 $\pm$ 1.1 \\	
 & CAS-AFQMC & $10^{-5}$ & 96317 & -462.88242 $\pm$ 0.00105 & -1636.0 $\pm$ 1.0 \\	 
\hline
\label{table:PD-631G}
\end{tabular}
\end{table*}

Building on the earlier discussion regarding the STO-6G basis set, our findings in Table~\ref{table:PD-631G} suggest consistent observations. Specifically, the disparity between RHF-AFQMC and MSD-AFQMC trials remains substantial for the PD structure using the 6-31G basis set. At $\epsilon = 10^{-3}$, LAS-AFQMC, utilizing roughly 1200 Slater determinants (SDs), aligns closely—within chemical accuracy—with CAS-AFQMC that encompasses at least 4700 SDs for the PD structure. The same observation is also found for $\epsilon = 10^{-5}$ where the number of determinant in the CAS trial is roughly 8 times larger than that of the LAS trial. For the PD structure utilizing the 6-31G basis set, we have observed unusual energy profiles in both LAS-AFQMC and CAS-AFQMC methodologies, where they initially produce comparable energy results. However, a significant decrease in energy readings is noticed after approximately 2000 blocks in both techniques. Using a reduced timestep of $\Delta\tau=0.001$ eliminates this unusual behavior, suggesting that it is a consequence of timestep errors. When calculating the average energy using only the data before this sudden drop (in the case of  $\Delta\tau=0.005$), there is a notable alignment in the average energies derived from the two different timesteps. It is important to note that this phenomenon is exclusive to the PD structure at the 6-31G basis set; such behavior is not observed in other structures we have studied. A detailed discussion on this observation is available in the Section S3 of the Supporting Information. Overall, the agreement between LAS-AFQMC and CAS-AFQMC is also observed for the PD5 structure across various cutoff values.

\begin{figure}[!ht]
    \centering
    \scalebox{0.35}{\includegraphics{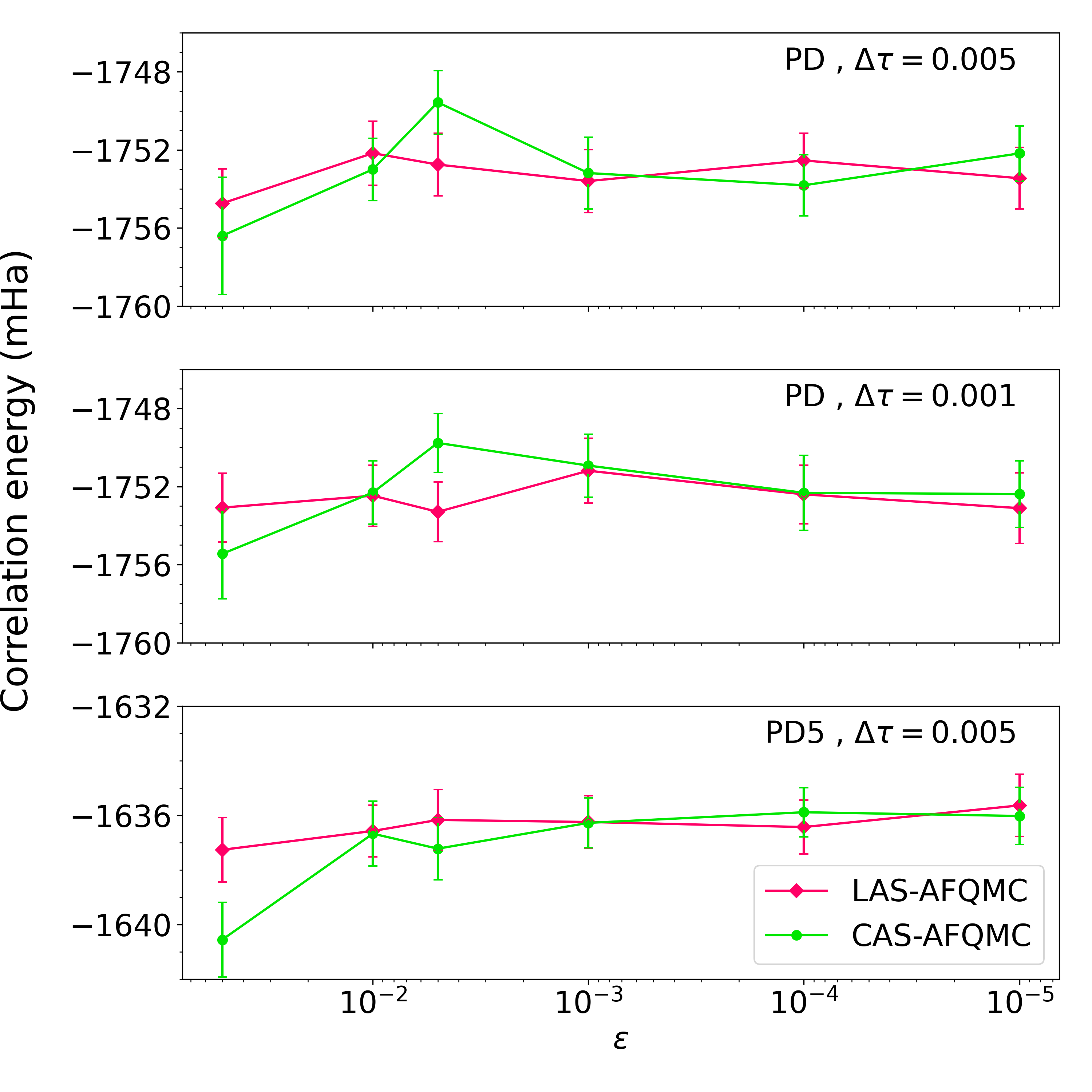}}
    \caption{Convergence of correlation energy for PD and PD5 plotted against $\epsilon$.}
    \label{fig:PD}
\end{figure}

For the  6-31G basis set, RHF-AFQMC recovers more correlation energy than either LASSCF or CASSCF. This contrasts with the STO-6G results, where RHF-AFQMC seemed to undercorrelate compared to LASSCF or CASSCF. This pattern further underscores the notion that minimal basis sets tend to amplify the effects of strong correlations. We delve deeper into the convergence of total energy concerning various cutoff values, as depicted in Figure~\ref{fig:PD}. For PD, achieving chemical accuracy in agreement between LAS-AFQMC and CAS-AFQMC necessitates a cutoff of at least $\epsilon=10^{-3}$. A similar observation is noted for PD5. In both configurations, LAS-AFQMC appears to converge more swiftly than CAS-AFQMC. This observation underscores the compact nature of the LAS trial. Furthermore, the benzene dimer, characterized by weak intermolecular coupling, emerges as an optimal candidate for DP-MSD-AFQMC in general and the LAS trial in particular. Collectively, our data on the benzene dimer serve as an ideal exemplar for our method. It showcases the potential to decompose multiple active spaces, separated by nonbonding interactions, into more manageable active spaces within the LASSCF framework. This approach offers a compact trial for AFQMC while preserving the precision required to capture both static and dynamic correlations in large molecular systems.

\section{Conclusions}
In this study, we proposed the use of direct-product trial states with ph-AFQMC to overcome computational challenges associated with studying strongly correlated molecules. This approach has the potential to be particularly effective for systems with weak- to non-coupling active subspaces such as those found in molecular crystals or van der Waals complexes in biological systems. By exploiting the inherent local nature of the multireference characters, we constructed compact wave functions that can be used as trial states in AFQMC simulations within the phaseless approximation.

Our numerical simulations of systems with varying electronic complexities demonstrate the potential of such solutions for studying much larger systems. However, in the case of systems with strong subspace coupling, like the \ce{C2H4N4} case study, our tailored solutions do not perform as well as the complete active space trial. We do not exclude the possibility that a clever choice of fragmentation could improve the overlap between the trial wave functions and the exact wave function, as observed in the \ce{H10} case study in our numerical simulations. However, such a solution would be more system-specific and would necessitate a systematic and meticulous examination based on chemical intuition. This limitation of our method necessitates further research on how to incorporate a small number of determinants that effectively capture the coupling between subspaces, avoiding unphysical decompositions in chemical systems while keeping the required number of determinants low, thereby preserving a low computational overhead. We emphasize that our approach can benefit from any advancements aimed at enhancing the scaling of ph-AFQMC, including algorithmic developments for local energy evaluation.\cite{Shee2017Jun,Shee2019Sep,Shee2019Apr}

While we demonstrated the use of LASSCF as an example of a direct-product wave function, our proposed technique can be applied to any method that allows for a decomposition of complete active spaces. Ongoing research is currently underway to generalize our implementation to other DP-MSD wave functions, such as ASD. It is important to note that our proposed method, which utilizes ph-AFQMC, achieves convergence to the exact energy as the trial states converge to the exact wave function, as demonstrated previously. This stands in contrast to conventional "static-then-dynamic" approaches, such as localized active space pair-density functional theory (LAS-PDFT),\cite{Pandharkar2021} which employs an energy correction on top of a LASSCF wave function that includes some static correlation but may not converge to the exact energy. On the other hand, our proposed method can converge to the exact energy as the DP-MSD trial states approach the exact wave function, as previously demonstrated in previous work.\cite{Lee2022}

Overall, our method demonstrates promise for conducting large-scale and accurate ph-AFQMC calculations on molecular systems with strong correlation and decomposable active space. This bears significant implications for predicting binding sites in proteins and other vital applications, including charge transport in DNA \cite{Eisinger1968,Gueron1967,Improta2016}. In future studies, we intend to further investigate the efficiency and accuracy of our approach in addressing these systems. Moreover, it would be beneficial to expand our method from non-coupling and weakly-coupling subspace systems to systems with strongly-coupling subspace in conjunction with the self-consistent optimization of the trial wave function.\cite{qin2023self} \\

\section*{Author Contributions} 

The manuscript was written through contributions of all authors. All authors have given approval to the final version of the manuscript. \\

\section*{Acknowledgements}
The authors would like to express their sincere gratitude to Changsu Cao, Yifei Huang for their valuable insights and engaging discussions. The authors extended their gratitude to Hang Li for his support and guidance.

\bibliography{refs}



\section*{Supplementary material}

# Scalable Quantum Monte Carlo with Direct-Product Trial Wave Functions

Hung Q. Pham,[*,†] Runsheng Ouyang,[‡] and Dingshun Lv[*,¶]

†ByteDance Research, San Jose, CA 95110, US.

‡ETH Zürich, Rämistrasse 101, 8092 Zürich, Switzerland.

¶ByteDance Research, Zhonghang Plaza, No. 43, North 3rd Ring West Road, 100098 Beijing, China.

E-mail: hung.pham@bytedance.com; lvdingshun@bytedance.com

# S1 Supplementary figures

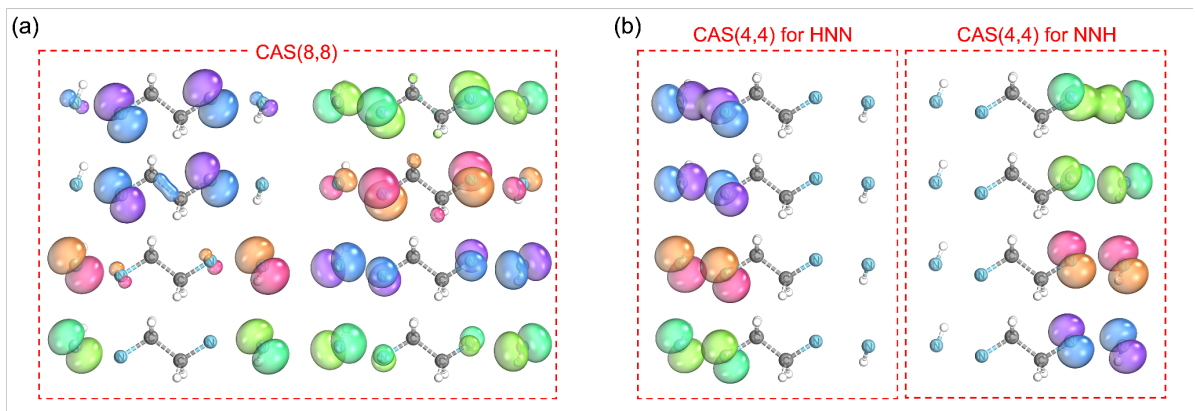

Figure S1: Orbitals used in CASSCF (a) and LASSCF calculation (b) for $C_2H_6N_4$

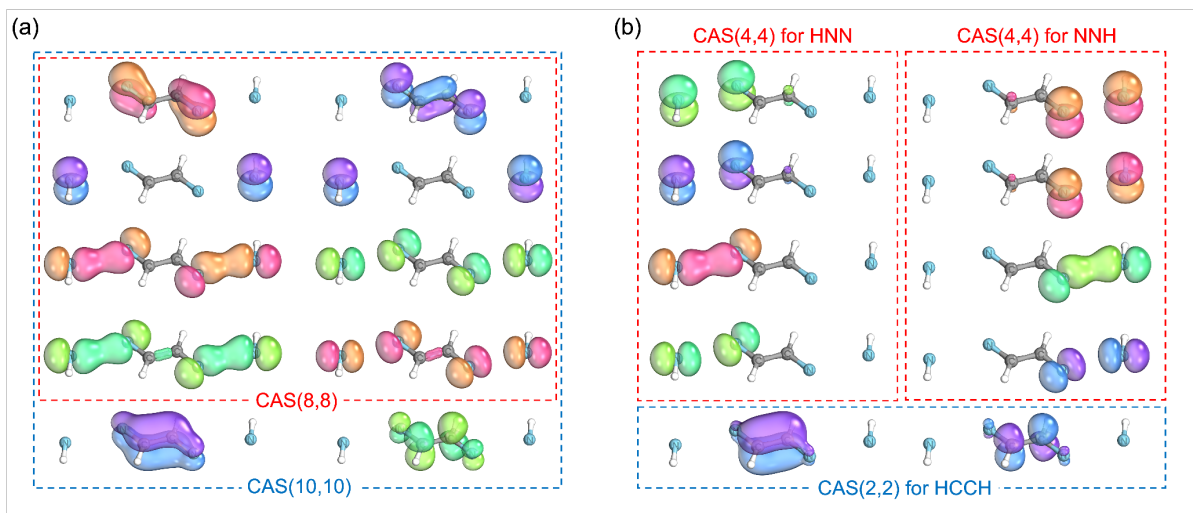

Figure S2: Orbitals used in CASSCF (a) and LASSCF calculation (b) for $C_2H_4N_4$

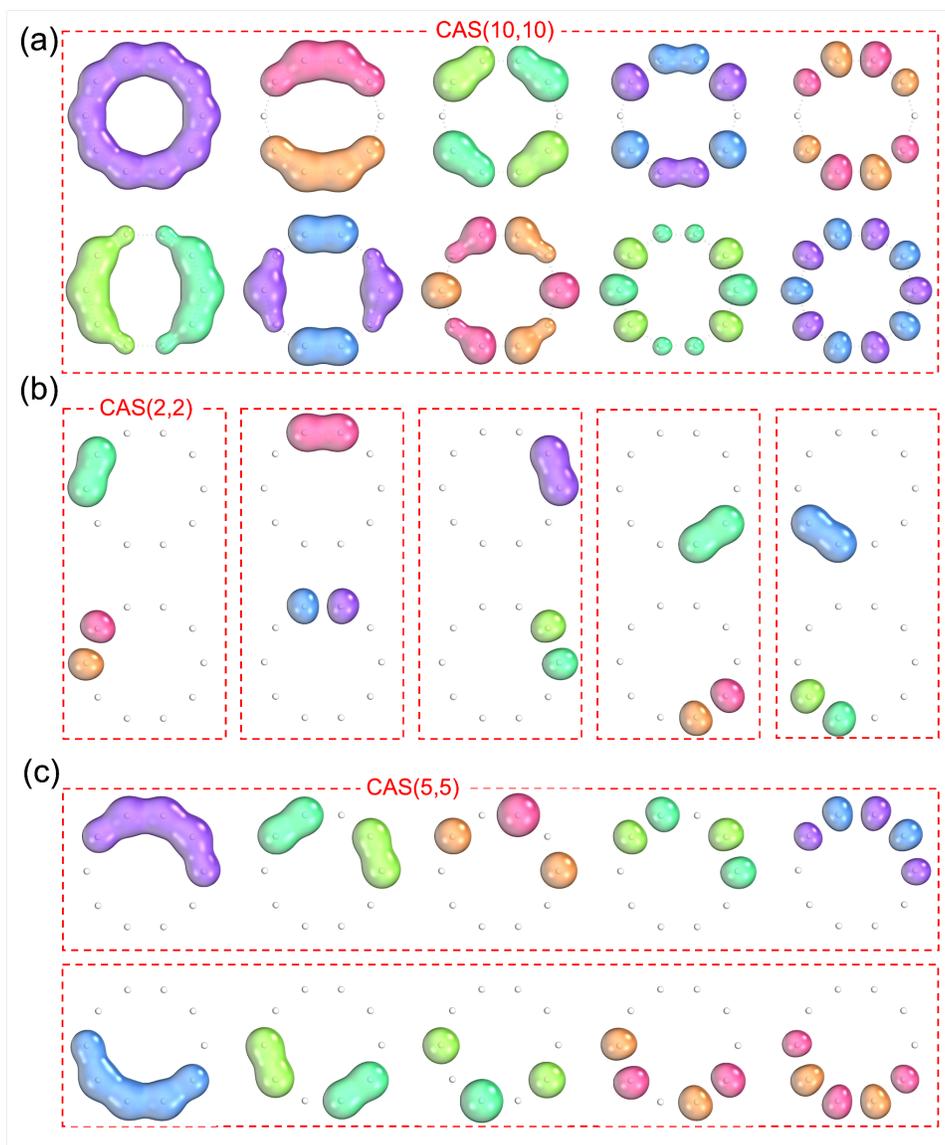

Figure S3: Orbitals used in CASSCF or LASSCF calculation for $H_{10}$: (a) CAS(10,10); (b) CAS(10,10) is partitioned into five CAS(2,2) active spaces; (c) CAS(10,10) is partitioned into five CAS(5,5) active spaces wich are anti-ferromagnetically coupled.

3

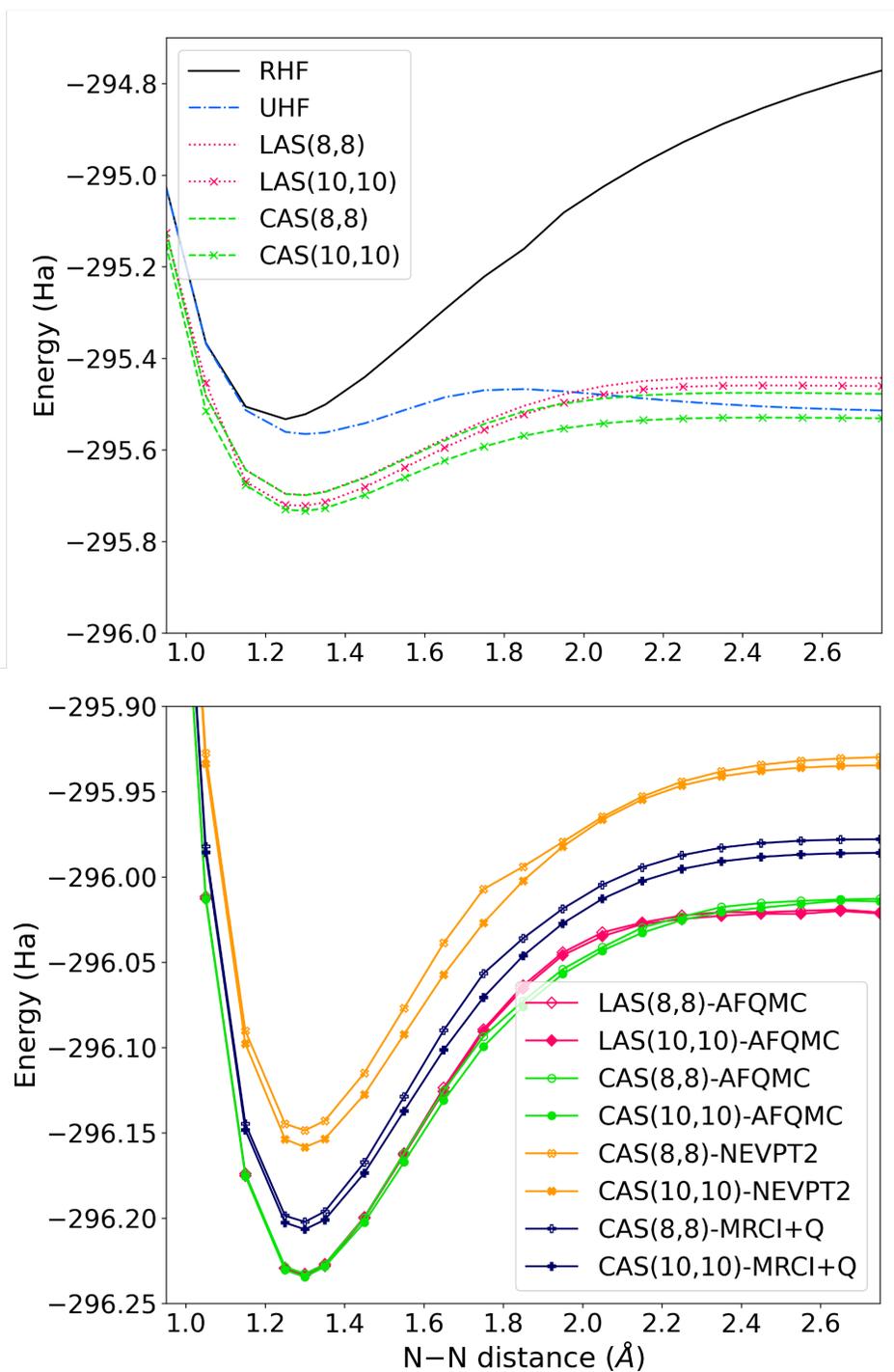

Figure S4: (a) Potential energy curves of $C_2H_4N_4$ obtained using LASSCF and CASSCF methods. (b) Potential energy curves of ph-AFQMC, NEVPT2, and MRCI+Q methods using different active spaces.

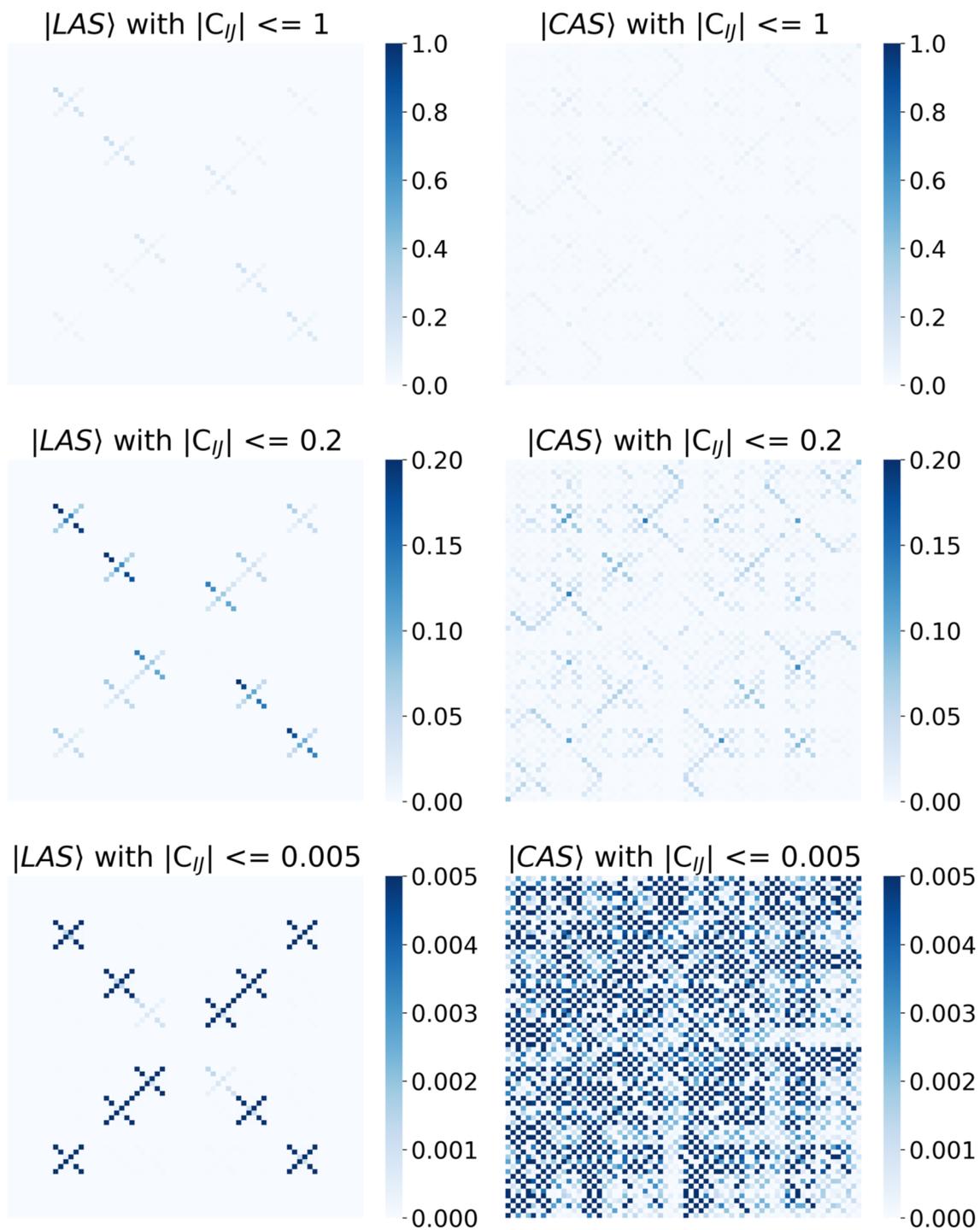

Figure S5: Visual heatmaps representing the CI tensors of LAS and CAS trials for $C_2H_6N_4$ across various threshold values.

## S2 Supplementary data

Table S1: Total energy of the parallel-displaced benzene dimer (PD) at varying cutoff values using the STO-6G basis set. The correlation energy is defined as the difference between ph-AFQMC and RHF, where the RHF energy is -460.25807861 Hartree.

| Method | $\epsilon$ | $N_c$ | $E \pm$ Error |
|---|---|---|---|
| RHF-AFQMC | | 1 | -461.12927 $\pm$ 0.00063 |
| LAS-AFQMC | $5 \times 10^{-2}$ | 21 | -462.08191 $\pm$ 0.00032 |
| | $1 \times 10^{-2}$ | 93 | -462.08232 $\pm$ 0.00026 |
| | $5 \times 10^{-3}$ | 229 | -462.08243 $\pm$ 0.00028 |
| | $1 \times 10^{-2}$ | 959 | -462.08289 $\pm$ 0.00062 |
| CAS-AFQMC | $5 \times 10^{-2}$ | 21 | -462.08191 $\pm$ 0.00036 |
| | $1 \times 10^{-2}$ | 190 | -462.08191 $\pm$ 0.00028 |
| | $5 \times 10^{-3}$ | 344 | -462.08190 $\pm$ 0.00028 |
| | $1 \times 10^{-3}$ | 3589 | -462.08260 $\pm$ 0.00067 |

Table S2: Total energy of the parallel-displaced benzene dimer (PD) at varying cutoff values using the 6-31G basis set. The correlation energy is defined as the difference between ph-AFQMC and RHF, where the RHF energy is -461.24260164 Hartree.

| Method | $\epsilon$ | $N_c$ | $E \pm$ Error | |
|---|---|---|---|---|
| | | | $\Delta\tau = 0.005$ | $\Delta\tau = 0.001$ |
| RHF-AFQMC | | 1 | -462.41604 $\pm$ 0.00060 | |
| LAS-AFQMC | $5 \times 10^{-2}$ | 31 | -462.99734 $\pm$ 0.00176 | -462.99568 $\pm$ 0.00175 |
| | $1 \times 10^{-2}$ | 125 | -462.99477 $\pm$ 0.00164 | -462.99507 $\pm$ 0.00156 |
| | $5 \times 10^{-3}$ | 319 | -462.99535 $\pm$ 0.00160 | -462.99590 $\pm$ 0.00152 |
| | $1 \times 10^{-3}$ | 1239 | -462.99620 $\pm$ 0.00162 | -462.99379 $\pm$ 0.00166 |
| | $1 \times 10^{-4}$ | 5415 | -462.99514 $\pm$ 0.00140 | -462.99500 $\pm$ 0.00150 |
| | $1 \times 10^{-5}$ | 12684 | -462.99605 $\pm$ 0.00157 | -462.99571 $\pm$ 0.00181 |
| CAS-AFQMC | $5 \times 10^{-2}$ | 29 | -462.99900 $\pm$ 0.00301 | -462.99805 $\pm$ 0.00230 |
| | $1 \times 10^{-2}$ | 207 | -462.99560 $\pm$ 0.00159 | -462.99491 $\pm$ 0.00162 |
| | $5 \times 10^{-3}$ | 480 | -462.99217 $\pm$ 0.00163 | -462.99237 $\pm$ 0.00151 |
| | $1 \times 10^{-3}$ | 4721 | -462.99579 $\pm$ 0.00183 | -462.99353 $\pm$ 0.00161 |
| | $1 \times 10^{-4}$ | 34057 | -462.99642 $\pm$ 0.00156 | -462.99492 $\pm$ 0.00192 |
| | $1 \times 10^{-5}$ | 106501 | -462.99478 $\pm$ 0.00141 | -462.99499 $\pm$ 0.00171 |

Table S3: Total energy of the 5 Å seperated parallel-displaced benzene dimer (PD5) at varying cutoff values using the STO-6G basis set. The correlation energy is defined as the difference between ph-AFQMC and RHF, where the RHF energy is -460.26007581 Hartree.

| Method | $\epsilon$ | $N_c$ | $E \pm$ Error |
|--------|-----------|-------|---------------|
| RHF-AFQMC | | 1 | -461.12843 ± 0.00060 |
| LAS-AFQMC | $5 \times 10^{-2}$ | 21 | -462.03547 ± 0.00039 |
| | $1 \times 10^{-2}$ | 93 | -462.03458 ± 0.00028 |
| | $5 \times 10^{-3}$ | 229 | -462.03468 ± 0.00027 |
| | $1 \times 10^{-3}$ | 959 | -462.03572 ± 0.00028 |
| CAS-AFQMC | $5 \times 10^{-2}$ | 25 | -462.03477 ± 0.00033 |
| | $1 \times 10^{-2}$ | 189 | -462.03484 ± 0.00028 |
| | $5 \times 10^{-3}$ | 335 | -462.03497 ± 0.00028 |
| | $1 \times 10^{-3}$ | 3533 | -462.03539 ± 0.00026 |

Table S4: Total energy of the 5 Å seperated parallel-displaced benzene dimer (PD5) at varying cutoff values using the 6-31G basis set. The correlation energy is defined as the difference between ph-AFQMC and RHF, where the RHF energy is -461.24639891 Hartree.

| Method | $\epsilon$ | $N_c$ | $E \pm$ Error |
|--------|-----------|-------|---------------|
| RHF-AFQMC | | 1 | -462.41398 ± 0.00062 |
| LAS | $5 \times 10^{-2}$ | 31 | -462.88366 ± 0.00118 |
| | $1 \times 10^{-2}$ | 125 | -462.88297 ± 0.00094 |
| | $5 \times 10^{-3}$ | 319 | -462.88257 ± 0.00111 |
| | $1 \times 10^{-3}$ | 1259 | -462.88264 ± 0.00096 |
| | $1 \times 10^{-4}$ | 4927 | -462.88283 ± 0.00099 |
| | $1 \times 10^{-5}$ | 9024 | -462.88204 ± 0.00114 |
| CAS | $5 \times 10^{-2}$ | 33 | -462.88695 ± 0.00137 |
| | $1 \times 10^{-2}$ | 211 | -462.88307 ± 0.00119 |
| | $5 \times 10^{-3}$ | 517 | -462.88362 ± 0.00114 |
| | $1 \times 10^{-3}$ | 4909 | -462.88267 ± 0.00091 |
| | $1 \times 10^{-4}$ | 33541 | -462.88229 ± 0.00090 |
| | $1 \times 10^{-5}$ | 96317 | -462.88242 ± 0.00105 |

Table S7: Total energy of PD calculated with the 6-31G basis set and an $\epsilon$ value of $5 \times 10^{-2}$

| | Number of Walkers | |
|--------|-------------------|------|
| | 640 | 1320 |
| LAS-AFQMC | -462.99734 ± 0.00176 | -462.99585 ± 0.00128 |
| CAS-AFQMC | -462.99900 ± 0.00301 | -462.99873 ± 0.00142 |

## S3 Energy profiles for the PD and PD5 structure at 6-31G basis set

Figure S6 shows the energy of the PD structure at the 6-31G basis set as a function of the number of blocks for LAS-AFQMC (pink) and CAS-AFQMC (green), across all cutoff values. We observed a bizarre drop in the energies for $\Delta\tau = 0.005$ a.u.. Using a smaller timestep of $\Delta\tau = 0.001$ a.u., this bizarre behavior disappears for both LAS-AFQMC abd CAS-AFQMC while the problem remains the same for CAS-AFQMC for a timestep of $\Delta\tau = 0.0025$ as shown in Figure S7. This strongly suggests that it is due to a time step error. Interestingly, the average energies obtained using the first 1800 blocks for $\Delta\tau = 0.005$ a.u. align well with the average energies obtained up to 6000 blocks for $\Delta\tau = 0.001$ a.u.. The first 500 and 1000 blocks are considered as equilibration and are discarded for the energy analysis of $\Delta\tau = 0.005$ a.u. and $\Delta\tau = 0.001$ a.u., respectively. We highlight that this behavior is exclusive to the PD structure at the 6-31G basis set and is not observed in other structures studied in this work. For example, for the PD5 structure, both timesteps agree very well, and there is no unphysical drop in the energies as shown in Figure S8.

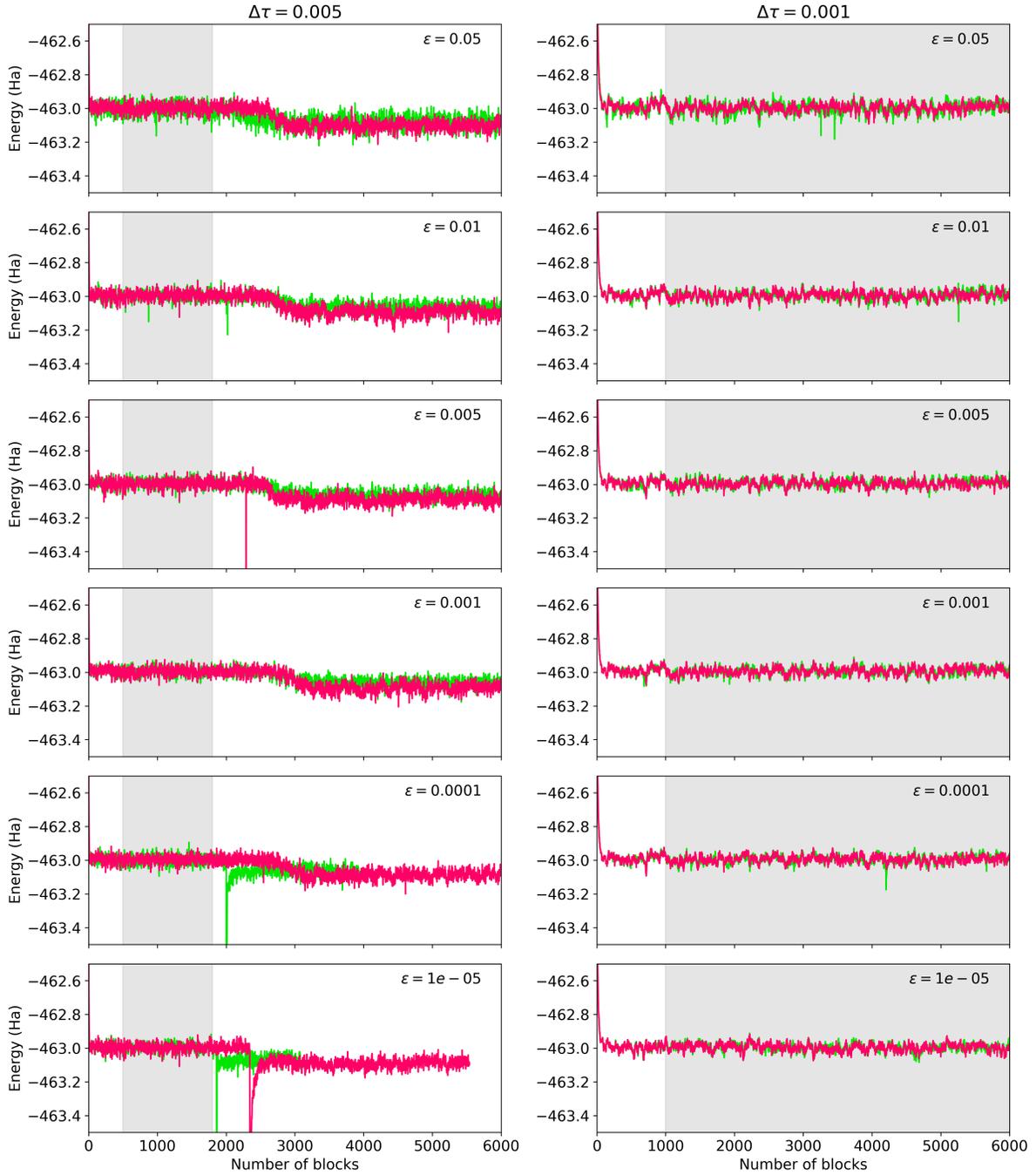

Figure S6: The energy of the PD structure at 6-31 basis set as a function of the number of blocks for LAS-AFQMC (pink) and CAS-AFQMC (green), across all cutoff values $\epsilon$ used in truncating the trial wave function. The left column corresponds to a time step of 0.001 a.u. $\Delta\tau = 0.001$ a.u., and the right column to a time step of 0.005 a.u. $\Delta\tau = 0.005$ a.u.. The shaded grey panel marks the blocks utilized for calculating the average energy and its variance, as detailed in Table S2.

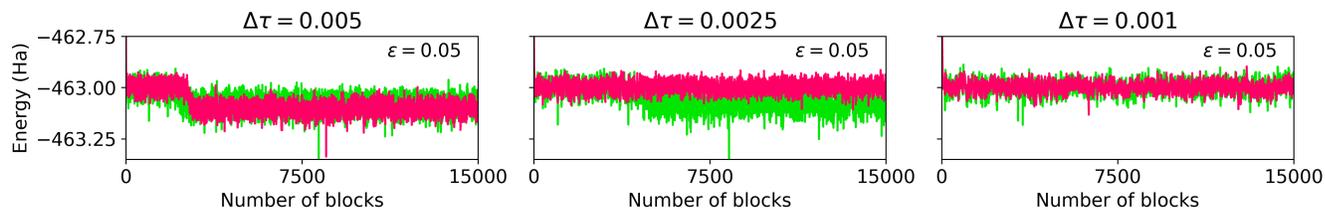

Figure S7: The energy of the PD structure at 6-31 basis set as a function of the number of blocks for LAS-AFQMC (pink) and CAS-AFQMC (green) at $\epsilon = 0.05$ and $\Delta\tau = 0.005, 0.0025, 0.001$ a.u. (from left to right).

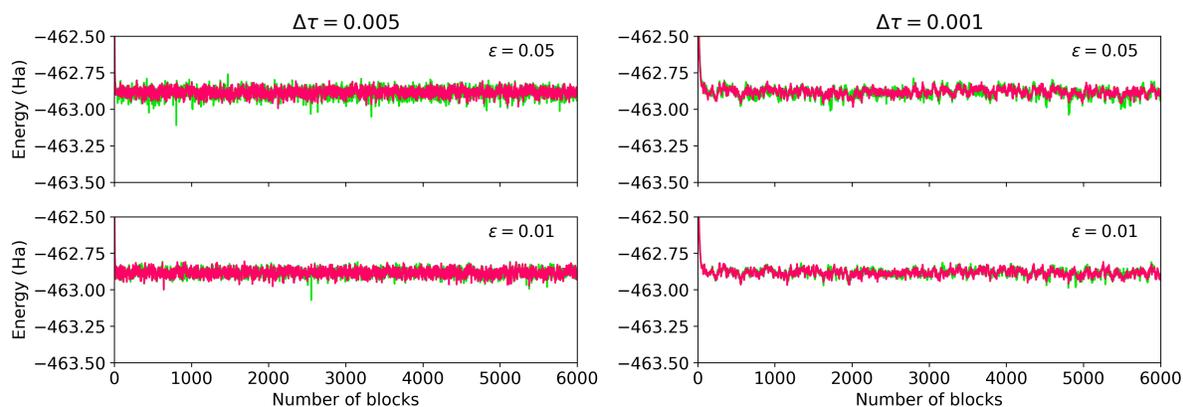

Figure S8: The energy of the PD5 structure at 6-31 basis set as a function of the number of blocks for LAS-AFQMC (pink) and CAS-AFQMC (green), across all cutoff values $\epsilon$ used in truncating the trial wave function. The left column corresponds to a time step of 0.001 a.u. $\Delta\tau = 0.001$ a.u., and the right column to a time step of 0.005 a.u. $\Delta\tau = 0.005$ a.u.



\end{document}